\documentclass[aps,showpacs,prd,amsmath,amssymb,floatfix,superscriptaddress]{revtex4}

\usepackage{epsfig}
\usepackage{graphicx,psfrag}
\usepackage{dcolumn}
\usepackage{bm}

\newcommand{\beq}{\begin{equation}}
\newcommand{\eeq}{\end{equation}}
\newcommand{\bea}{\begin{eqnarray}}
\newcommand{\eea}{\end{eqnarray}}
\newcommand{\hf} {\frac{1}{2}}
\newcommand{\nonu}{\nonumber\\}
\newcommand{\nn}{\nonumber\\}

\newcommand{\sla}[1]{\mbox{$#1\!\!\!/$}}

\newcommand\eqn[1]     {Eq.\,(\ref{#1})}
\newcommand\eqns[2]    {Eqs.\,(\ref{#1}) and~(\ref{#2})}

\newcommand\fig[1]     {Fig.\,{\ref{#1}}}
\newcommand\sect[1]    {Sect.\,{\ref{#1}}}

\newcommand\app[1]     {Appendix~\ref{#1}}

\def\ord#1{{\cal O}(#1)}

\def\mr#1{{\mathrm{#1}}}
\def\ci{{\rm i}}

\def\t{\tilde}

\begin{document}

\title{Massless fermions in multi-flavor QED$_2$}

\author{S. Nagy}

\affiliation{Department of Theoretical Physics, University of Debrecen, Debrecen, Hungary}

\begin{abstract}
The ground state of the multi-flavor 2-dimensional quantum electrodynamics (QED$_2$)
is determined in the presence of finite baryon density, and it is shown that the model possesses
two phases, the low density phase where the external baryon density is totally screened, and the
high density phase, where the screening is only partial. The renormalization
of the bosonized version of the model is also performed for both the zero and the finite density model
giving massless multi-flavor QED$_2$ in both cases.
\end{abstract}


\maketitle

\section{Introduction}

The phase structure and the confinement mechanism of non-abelian models are usually
investigated in much simpler usually two dimensional 'toy' models \cite{Coleman,Fischler}.
These models are sometimes analytically solvable, e.g. the massless QED$_2$ \cite{Schwinger}
(usually referred to as Schwinger model) shows the chiral condensate and a mass gap
which are supposed to be essential elements in modern physics. The bosonized version of
the Schwinger model is a free theory, which turns into interacting when the fermionic mass
is nonzero \cite{Adam}. QED$_2$ possesses a phase transition at $m/g_c\sim 0.31$ as was
shown by density matrix renormalization group (RG) technique \cite{dm_rg} or by continous
RG method \cite{Nagy_MSG}. The critical value of $m/g_c$ separates the large coupling
($g \gg m$) phase with a unique vacuum characterized by the field variable $\phi=0$, and
the weak coupling ($g \ll m$) phase where due to spontaneously broken reflection symmetry
the model has non-trivial vacua, at around $\phi = \pm \sqrt{\pi}/2$.

The ground state of quantum chromodynamics (QCD) is a color superconductor at high densities
\cite{QCDbcs} and a periodic chiral condensate appears in the coordinate space at large number of colors
$N_c$ \cite{Deryagin} which was recapitulated in the framework of renormalizaton group (RG)
equations by mapping QCD into a Thirring type model \cite{Shuster}. The high density
behavior of non-abelian models are investigated in the framework of toy models too \cite{Thies}.
The Schwinger model remains exactly solvable in the presence of an external
finite charge density and also shows periodic chiral condensate \cite{Kao,Metlitski}.
In our previous work \cite{Nagy_schw} we showed that in the finite density QED$_2$ the system
exhibits a single periodic phase in the thermodynamic limit for arbitrary charge densities,
furthermore the periodic structure built up in the ground state has decreasing amplitude and
wavelength with increasing charge density. It is assumed \cite{Fischler} that a phase
transition appears in the 2-flavor QED$_2$ as the density is increased. By the bosonization
technique \cite{bos} both the massless and the  multi-flavor QED$_2$ can be converted to such local
scalar field theoretical models which contain periodic self-interaction potentials.
The bose form of the multi-flavor QED$_2$ is a sine-Gordon--type scalar field theory
\cite{Hosotani,NQED2bos,Nandori_PLB}, where periodic self-interactions are described by
2-dimensional sine-Gordon (SG) fields which coupled by a mass matrix.
This coupled SG (or layered sine-Gordon (LSG)) model can also be used
to describe the vortex dynamics of magnetically coupled layered superconductors \cite{LSG},
where the number of flavors in the high energy model is identical to the number of layers
of the condensed matter system \cite{Nandori_PLB}.
Moreover coupled SG type models have been used to investigate the vortex dynamics
of Josephson coupled superconductors \cite{Benfatto}.

The phase transition of these models was obtained from the microscopic theory which is
formulated in the high energy or ultraviolet (UV) region. In order to obtain the low
energy/infrared (IR) physics, where the measurements are performed and the
quantum fluctuations are taken into account we need renormalization.
The original, fermionic models so as the toy models contain strong couplings and it disables performing a
perturbative renormalization, and it makes rather difficult to develop a functional
renormalization group (RG) method in the fermionic models since the evolution should be
started from a perturbative region where the theory is almost interaction free. Furthermore
the RG equations has to preserve the gauge symmetry \cite{gauge-RG}. However the bosonized
version of the toy models can be easily treated by the functional RG method
\cite{Nandori_PLB,Nagy_MSG,Nagy_SG}.

Our goal in this article to clarify the phase transition of the finite density 2-flavor
QED$_2$ by the methods used in \cite{Nagy_schw}. We calculate numerically (and also
analytically in the low and high density limits) the ground state
field configurations for finite external baryon density in the tree-level.
The results show that an induced baryon density appears which screens the external one
totally (partially) for low (high) external baryon densities respectively, showing the existence of
two phases as is conjectured in \cite{Fischler}. We also perform the RG procedure to determine
the IR physics of the zero and finite density 2-flavor QED$_2$ and then generalize
the results for arbitrary number of flavors. We choose the so-called Wegner-Houghton (WH)
RG method \cite{WH-RG} in order to obtain the blocked potential for the model, which uses
a gliding sharp cut-off $k$. The external baryon density produces coordinate dependent
non-trivial saddle points, therefore we use the tree-level blocking relation
\cite{Polonyi_tree,Nagy_MSG,Nagy_SG} to get the blocked interaction potential. The RG
evolution gives the flow of the couplings for the LSG model from which the evolution of
the couplings of the fermionic model can be easily determined.

The paper is organized as follows. In \sect{sec:model} we show the connection
between the fermionic and the bosonized models and then in \sect{sec:phase}
we determine the ground state field configuration of the finite density 2-flavor QED$_2$
and map its phase structure. We derive the evolution of the coupling in the framework of the
WH-RG method in \sect{sec:ren}, and determine the flow of the couplings
in the case zero, low and high densities. Finally, in \sect{sec:con} the conclusion is drawn up.

\section{The model}\label{sec:model}
The multi-flavor QED$_2$ containing $N$ Dirac fields with identical fermionic charge $e$ and mass $m$
has the Lagrangian density
\beq
{\cal L}=-\frac14 F_{\mu\nu}F^{\mu\nu}+\sum_{i=1}^N\bar\psi_i\gamma^\mu(\partial_\mu-
\ci e A_\mu)\psi_i-m\sum_{i=1}^N\bar\psi_i\psi_i,
\eeq
where $F_{01} = \partial_0 A_1-\partial_1 A_0$. One can turn from fermionic field variables
$\bar\psi_i,\psi_i$ into bosonic ones $\phi_j$ by the bosonization rules \cite{Coleman 1976,Fischler}
\bea
:\bar\psi_i\psi_i: &\to& -cmM\cos(2\sqrt{\pi}\phi_i) \nonu
:\bar\psi_i\gamma_5\psi_i: &\to& -cmM\sin(2\sqrt{\pi}\phi_i) \nonu
:\bar\psi_i\gamma_\mu\psi_i: &\to&
 \frac1{\sqrt{\pi}}\varepsilon_{\mu\nu}\partial^\nu \phi_i \nonu
:\bar\psi_i\ci \sla\partial\psi_i: &\to& \frac12 N_m (\partial_\mu\phi_i)^2,
\eea
where $N_m$ means normal ordering with respect to the fermion mass $m$, $M=e/\sqrt{\pi}$, and
$c=\exp(\gamma)/2\pi$, with the Euler constant $\gamma=0.5774$.
The presence of a non-vanishing external or background densities does not affect these transformation
rules \cite{Kao}. The Hamiltonian of the system in Coulomb gauge is given by
\beq
{\cal H} = \sum_{i=1}^N\int_x \bar\psi_i(x)(\ci \gamma_1\partial_1+m)\psi_i(x)
-\frac{e^2}{4}\int_{x,y} j_{0,x}|x-y|j_{0,y},
\label{hamcoul}
\eeq
with $\int_x=\int_0^T dx^0 \int_{-L}^L dx^1$ and
\beq
j_{0,x} = :\sum_{i=1}^N \bar\psi_i(x)\psi_i(x): =\frac1{\sqrt{\pi}}\partial_1\sum_{i=1}^N\phi_i(x).
\eeq
The resulting bosonized form of the Hamiltonian is
\beq
{\cal H} = N_m\int_x \left[\frac12\sum_{i=1}^N\Pi^2_i(x)+\frac12\sum_{i=1}^N(\partial_1\phi_i(x))^2
+\frac{e^2}{2\pi}\left(\sum_{i=1}^N \phi_i(x)\right)^2
-cm^2\sum_{i=1}^N \cos\left(2\sqrt\pi\phi_i(x)\right)\right],
\label{hamsc}
\eeq
where $\Pi_i(x)$ denotes the momentum variable canonically conjugated to $\phi_i(x)$.
Let us assume that the two distinct flavored fermion has the same mass $m$ but opposite
charge $e$. The resulting bosonic Hamiltonian corresponds to the LSG model
\beq
{\cal H} = N_M\int_x \left[\hf\Pi^2_1+\hf\Pi^2_2
+\hf(\partial_1\phi_1)^2+\hf(\partial_1\phi_2)^2+\frac{M^2}2(\phi_1-\phi_2)^2
-u(\cos(2\sqrt\pi\phi_1)+\cos(2\sqrt\pi\phi_2))\right]
\label{hamlsg}
\eeq
with two layers ($N=2$), and $u=cme/\sqrt{\pi}$.
The assumption for the charges corresponds to the
situation where the net electric charge is zero which means that the matter is colorless.
It is assumed that this kind of colorless matter might exist in nuclear stars. Let us denote
the two types of bose field as $\phi_i$ with $i=1,2$. After introducing the new fields as
\beq
\phi_\pm = \frac1{\sqrt{2}}(\phi_1\pm \phi_2),
\eeq
the charge density $j_0\equiv\rho$ and the baryon density $B$ can be written as
\beq
\rho = \frac2{\sqrt{\pi}}\partial_1\phi_-,\quad B = \frac2{\sqrt{\pi}}\partial_1\phi_+.
\eeq
An external baryon charge density can be introduced by replacing $\rho$ by
$\rho+\rho_0$. If we define the corresponding classical potential as
\beq
\rho_0 = \frac2{\sqrt{\pi}}\partial_1\phi_c.
\eeq
then the uniform, constant external baryon density is simply $\phi_c=bx$.
We separate this linear term in the space direction $\phi_+=\tilde \phi_++bx$,
introduce the chemical potential $\mu$ and assume that it is non-vanishing on the
interval $[-L;L]$. Taking $b=-2\mu2/\sqrt{\pi}$ then Hamiltonian reduces to
\beq\label{hamlay}
{\cal H} = N_M\int_x\left[\hf \pi^2_-+\hf(\partial_1\phi_-)^2+M^2\phi^2_-
+\hf \pi^2_+ +\hf(\partial_1\tilde\phi_+)^2
-u'\cos(\sqrt{2\pi}\phi_-)\cos(\sqrt{2\pi}(\tilde\phi_+ +bx))\right],
\eeq
due to the normal ordering with respect to the bosonic mass $M$ where the dimensionful coupling $u'$
\beq
u' = c' m^{3/2}e^{1/2}
\eeq
is introduced, and $c'=c(2/\pi)^{1/4}$. However, the periodic part of the Hamiltonian
also gives a contribution to the mass spectrum, i.e. the normal ordering can also be defined
with respect to the 'total' mass. In this case the relation between the mass gap of the
multi-flavor bosonic model and the parameters of the original fermionic theory is
\beq
M_{gap} = 2.008 \cdot m^{\frac{N}{N+1}}e^{\frac1{N+1}}
\eeq
where $N$ is the number of flavors \cite{NQED2bos}. The Hamiltonian (\ref{hamlay}) is a functional of the
field configurations $\phi_-$ and $\phi_+$, and it gives the energy of the system on the tree level.
We look for the ground state field configuration of the model as the function of the density $b$.

\section{The tree level phase structure}\label{sec:phase}

\begin{figure}[ht]
\includegraphics[width=8cm]{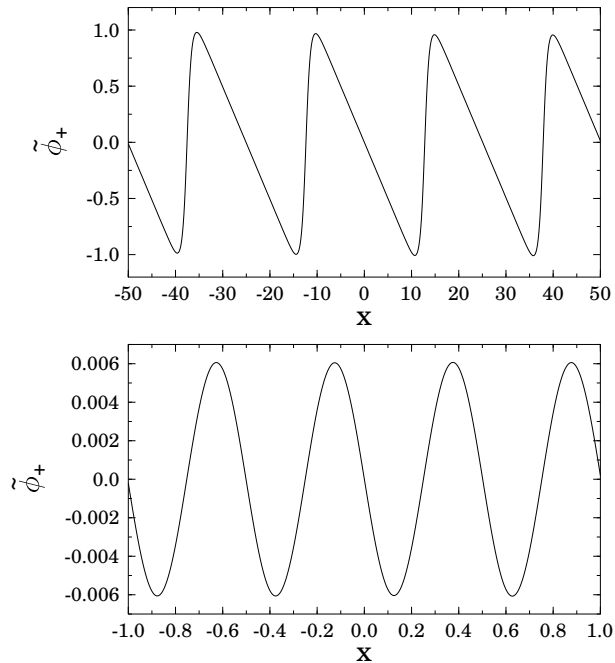}
\caption{The form of the field configuration for $b=0.1$ (upper figure)
and $b=5$ (lower figure).
\label{fig:phi}}
\end{figure}

\begin{figure}[ht]
\includegraphics[width=8cm]{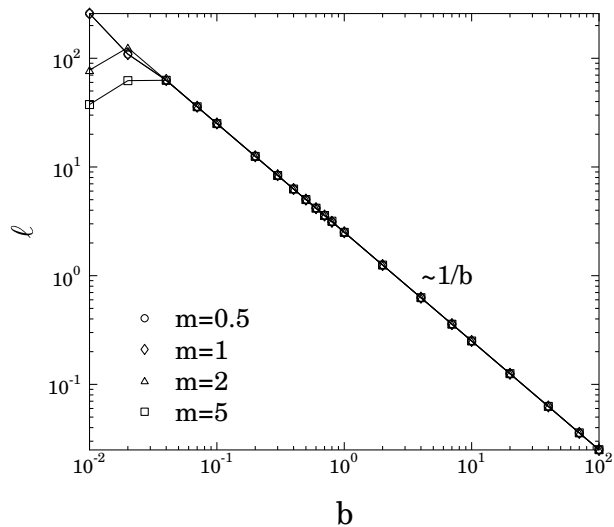}
\caption{The baryon density dependence of the wavelength $\ell$ of the field configuration $\tilde\phi_+$.
\label{fig:l}}
\end{figure}

\begin{figure}[ht]
\includegraphics[width=8cm]{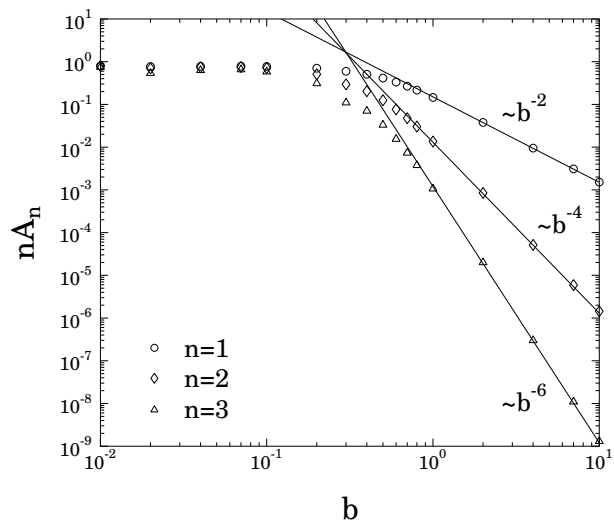}
\caption{The baryon density dependence of the Fourier modes $A_n$, $n=1,2,3$. $A_n$ follow a
power law behavior at high densities as it was obtained analytically.
\label{fig:fourier}}
\end{figure}

\begin{figure}[ht]
\includegraphics[width=8cm]{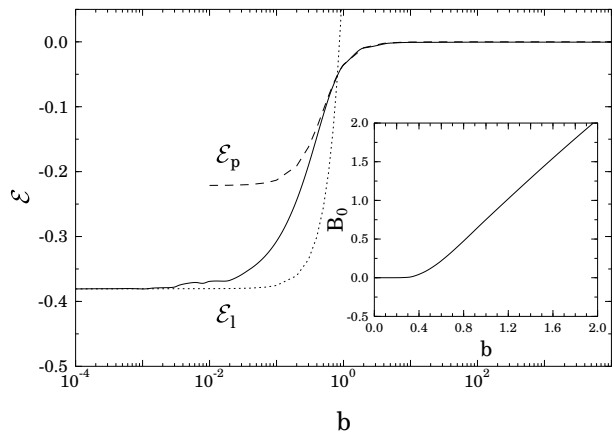}
\caption{The baryon density dependence of the energy density in case of linear ${\cal E}_l$ (dotted line)
and periodic ${\cal E}_p$ (dashed line) configurations. We also plotted the numerically determined
total energy density by a solid line. At low densities the linear, at high densities
the periodic field configuration coincides with the numerical results.
The inset shows how the baryon density $B_0$ increases as the background baryon density $b$ grows.
\label{fig:en}}
\end{figure}

The minimum of the energy is searched numerically among the static field configurations,
$\pi_\pm(x)=0$ by means of conjugate gradient method as a function of the finite baryon
density $b$. We made numerical calculations in order to minimize the tree level energy as
the functional of the field variables $\phi_-$ and $\phi_+$. The results show that at the energy
minimum $\phi_-(x) =0$ for all value of $b$. This result is not surprising since in the
Hamiltonian in \eqn{hamlay} one has a massive sine-Gordon (MSG) model
\cite{Coleman,Fischler,Coleman 1976,MSG} with zero density for the field variable
$\phi_-(x)$, and this model exhibits trivial field configurations on the tree level approximation
\cite{Nagy_schw}. Therefore we  can treat a simpler Hamiltonian
\beq\label{hamsgb}
{\cal H} = \int_x\left[\hf \pi^2_+ +\hf(\partial_1\tilde\phi_+)^2
-u'\cos(\sqrt{2\pi}(\tilde\phi_+ +bx))\right],
\eeq
which corresponds to a SG model \cite{SG} with finite density.
\eqn{hamsgb} leads to the equation of motion
\beq\label{eqmot}
\left(\partial_0^2-\partial_1^2\right)\tilde\phi_++\sqrt{2\pi}u'\sin(\sqrt{2\pi}(\tilde\phi_+ +bx))=0.
\eeq
Considering the static equation of motion and using a simple redefinition of the field variable
$\sqrt{2\pi}(\tilde\phi_++bx)\to \tilde\phi_+$ gives the static SG equation \cite{kink} which is
identical to the evolution equation of a pendulum. Depending
on the initial energy $C$ of the pendulum, the model has two phases. When $C$ is large
then it makes periodic rotation corresponding to the kink (or antikink) crystal in the
original model \cite{kink}. In the low energy phase the pendulum swings, and we have a kink-antikink (KA)
crystal. The external baryon charge contribute to the energy by the term $\pi b^2$. Then the large
density limit corresponds the KA crystal, while at low densities the kink crystal solution appears.
The general analytic solution is
\beq
\tilde \phi_+ = -bx+\frac1{\sqrt{2\pi}}\mr{am}\left(\frac{\sqrt{2\pi u'}x}{k},r^2\right)
\eeq
with $\mr{am}(x,r^2)$ the Jacobian elliptic function and $r=\sqrt{2/(2+C/2\pi u')}$.
The coupling $u'$ plays key role in the RG procedure, so we keep it explicitly.
We consider the limiting low density ($r\to\infty$ or small values of $b$) and high density
($r\to 0$ or high values of $b$) cases.
The series representation of the function $\mr{am}(x,r^2)$ is
\bea\label{am}
\mr{am}(x,r^2) &=& \frac{\pi x}{2K(r)}+2\sum_{n=1}^\infty\frac{q(r)^n}{n(q(r)^{2n}+1)}
\sin\frac{n\pi x}{K(r)},
\eea
where $K(r)$ is the complete elliptic integral of the first kind and $q(r)=\exp(-\pi K(1-r)/K(r))$.
In the low density limit $q(r)\to -1$ and the second term is just the Fourier expansion of the first term
with opposite sign in the r.h.s. of \eqn{am}, giving $\mr{am}(x,r^2) \to 0$, so
\beq\label{lowtpi}
\tilde\phi_+ =-bx
\eeq
as is conjectured in \cite{Fischler}.
It gives a constant induced density equal to the external one with opposite sign,
resulting total screening.
At large densities the field configuration changes. When $b$ is large $C\approx\pi b^2$ and
the amplitude $A_n$ of the $n$th Fourier expansion in \eqn{am} scales as $A_n\sim 1/b^{2n}$ therefore
it is a good approximation to keep only the fundamental mode. Then the field configuration becomes
\beq\label{hightpi}
\tilde\phi_+ =-\frac{u'}{\sqrt{2\pi}b^2}\sin(\sqrt{2\pi}bx),
\eeq
a sinusoidal type one, giving sinusoidal induced density which only partially
screens the external density, and the net baryon density turns to finite values signalling
a phase transition going from low densities to high ones.
In \fig{fig:phi} two typical numerically obtained field configurations are plotted.
The wavelength $\ell$ is insensitive for the fermion mass $m$,
furthermore it decreases according to a power law in $b$ for several orders of the density.
The numerical fit give that $\ell=\sqrt{2\pi}/b$ equal to the `specific volume' of the
external density similarly to the case of MSG model \cite{Nagy_schw}, as can be seen in \fig{fig:l}.
The amplitude $A$ is a constant and independent of fermion mass for low densities since
it is independent of the coupling $u'$. Due to the linear form
of $\t \phi_+$ its value is $A=b\ell/2=\sqrt{\pi/2}$. Going to
high density regime the amplitude decreases as $A\sim 1/b^2$ and an $m$ dependence appears.
We plotted the amplitudes $A_n$ in \fig{fig:fourier}.
The limiting low and high density cases, is in a good agreement with its analytically calculated
value in \eqns{lowtpi}{hightpi}. \fig{fig:fourier} shows that the phase transition appears
around the critical value $b_c$.
Using the field configurations obtained in \eqns{lowtpi}{hightpi} one can calculate the
total energy density ${\cal E}$ by inserting them back into \eqn{hamsgb}.
For low densities it is ($l$ refers to $l$inear field configurations)
\beq
{\cal E}_l = \frac{b^2}2-u'.
\eeq
while in high densities one obtains ($p$ refers to $p$eriodic field configurations)
\beq\label{Eper}
{\cal E}_p=-\frac{u'^2}{4b^2}
\eeq
which shows that the energy density is always negative and the field
configuration in \eqn{hightpi} is energetically favourable in comparison with a trivial one
which would give ${\cal E} = 0$.
In \fig{fig:en} we plotted ${\cal E}_p$ and ${\cal E}_l$ and the numerically determined
total energy density. At low densities
the linear field configuration is preferable and for high densities the
sinusoidal one.
The high density phase then contains a coordinate dependent ground state which is
supposed to flatten out when one takes into account the quantum fluctuations beyond
the tree level approximation \cite{Fischler}.
The critical density $b_c$ which separates the low and the high density
phases can be read off correctly from \fig{fig:fourier} as the intersection of the Fourier modes,
and it is $b_c\approx 0.3$.
The baryon density at the origin $B(x{=}0)\equiv B_0$ can play the role of the order parameter of the
phase transition.
It is zero in the low density phase and is finite in the high one, as can bee seen in the
inset of \fig{fig:en}. According to the inset of \fig{fig:en} the phase transition seems a continous one.

\section{Renormalization}\label{sec:ren}

The phase structure of the finite density 2-flavor QED$_2$ is determined at the tree level
where the observational scale $k$ is in the UV regime. We use the RG method to lower the
scale $k$ into the IR limit in order to consider the effect of the quantum fluctuations
systematically. Here we perform the RG treatment of the LSG model for zero, and then
for finite density. We treat the model in a more general form containing the arbitrary frequency
parameter $\beta$. The bosonization works for the specific choice $\beta^2=4\pi$ and gives the
Hamiltonian in \eqn{hamlsg}, and in this article we map the phase structure as the function of the
baryon density $b$. However the phase structure can be also considered in terms of the frequency
$\beta$ with the critical value $\beta_c^2=16\pi$ in the LSG model \cite{Nandori_NPB}.
At the scale where the microscopic theory is formulated
the local potential approximation (LPA) is usually valid, implying that during the blocking
transformation a constant field configuration gives the minimum of the potential \cite{WH-RG,diff},
furthermore the RG methods based on the evolution of the effective action \cite{effac}
are also formulated with constant field configurations.
We showed in the preceding section, that for the finite-density 2-flavor QED$_2$ it is not
the case. The non-trivial saddle point field configuration makes unstable
modes, and one should take into account coordinate dependent field configuration
to find the real extremum of the Lagrangian, or at least better than the constant
field configuration could give \cite{Polonyi_tree,Wett_spin}. Therefore we choose
the WH-RG method where one can change to tree level blocking relation when
the non-trivial saddle point appears \cite{Polonyi_tree}. In this RG method
the blocked potential can be derived by means of the differential RG in momentum space at the
gliding momentum scale $k\in [0,\Lambda]$. In the WH-RG approach the field
fluctuations are ordered by their decreasing frequencies and during the blocking step
one integrates out the high-frequency modes above the cut-off $k$, keeping the generating
functional invariant. We decompose the field variables into the high-frequency
$\phi^\prime_x= \int_{|p|>k} \phi_p e^{ipx}$ and the low-frequency
$\phi_x= \int_{0\le|p|\le k} \phi_p e^{ipx}$ modes. The high-frequency modes are integrated
out step by step in infinitesimal momentum shells of thickness $\Delta k$,
\beq
e^{-S_{k-\Delta k}[\phi]}=\int{\cal D}[\phi']e^{-S_k[\phi+\phi']},
\label{RG}
\eeq
with $S_k[\phi]$ the blocked action in Euclidean spacetime.
The higher-frequency Fourier modes are split further into the sum of
the saddle point field configuration $\phi^{sp}$ and the remaining field
fluctuations: $\phi'=\phi^{sp}+\varphi$. In order to evaluate the blocked action in
\eqn{RG}, we expand the blocked action in Taylor series around its saddle point field
configuration and then, for trivial saddle points $\phi^{sp}=0$ one arrives at the WH-RG equation
\cite{Nandori_NPB}
\bea
\label{lsgwh}
\left(2 + k\partial_k \right) {\tilde V}_k
& =& - \frac1{4\pi} \log\left(
(1 +  {\tilde V}^{11}_k)(1 + {\tilde V}^{22}_k) - ( {\tilde V}^{12}_k)^2 \right),
\eea
for the LSG model with $N=2$ and the second derivatives
${\tilde V}^{ij}_k=\partial_{\phi_i}\partial_{\phi_j}{\tilde V}_k$
(all dimensionless quantities are denoted by a tilde superscript), which is introduced
as the sum of the dimensionless mass term and the dimensionless periodic potential,
\beq
\label{lsgV}
{\tilde V}_k (\phi_1, \phi_2)= \tilde M(\phi_1 -\phi_2)^2
+ {\tilde U}_k(\phi_1, \phi_2),
\eeq
which is the generalized Euclidean form of the Hamiltonian in \eqn{hamlsg}.
The ansatz for the periodic part of the potential is
\beq\label{lsgU}
\tilde U_k = \sum_{n_1,n_2}\left[\tilde u_{n_1n_2}\cos(n_1\beta\phi_1)\cos(n_2\beta\phi_2)
+\tilde v_{n_1n_2}\sin(n_1\beta\phi_1)\sin(n_2\beta\phi_2)\right].
\eeq
The scale-dependence is entirely encoded in the dimensionless couplings $\tilde u_{n_1n_2}=u_{n_1n_2}/k^2$
and $\tilde v_{n_1n_2}=v_{n_1n_2}/k^2$. Inserting \eqn{lsgV} into the functional evolution
equation in \eqn{lsgwh} and Fourier expanding it
we obtain a set of coupled differential equations for these couplings.
The right hand side of the WH-RG equation in \eqn{lsgwh} turns out to 
be periodic, while the left hand side contains periodic and non-periodic 
parts, as well. Separating them, we obtain a trivial evolution,
\beq
\tilde M_k = \tilde M_\Lambda\left(\frac{k}{\Lambda}\right)^{-2},
\eeq
hence the dimensionful bosonic mass $M_k$ remains constant during the RG procedure.
We note that the RG flow equation in \eqn{lsgwh} keeps the periodicity of the periodic
piece  $\tilde U_k$ of the blocked potential in both directions of the
internal space with unaltered length of periods, therefore $\beta$ does not evolve in LPA.

During the successive integrations of the higher-momentum modes of 
\eqn{lsgwh} by the method of steepest descent, it is assumed that the 
saddle points ($\phi^{sp}_1$, $\phi^{sp}_2$) are zero. It remains trivial till the
second functional derivative of the blocked action $S''_k$ (the argument of the
logarithm in \eqn{lsgwh}) is positive definite. When
\beq\label{sicond}
S''_k=(1+\t V^{11}_k)(1+\t V^{22}_k) - (\t V^{12}_k)^2)=0,
\eeq
the restoring force for the field fluctuations
with momenta in the momentum shell $k-\Delta k <|p|<k$ vanish, i.e. their amplitudes can
grow to finite values, and $\phi^{sp}$ becomes non-vanishing, and the spinodal instability
appears. Then, the tree level blocking relation is needed \cite{Polonyi_tree,Nagy_MSG,Nagy_SG,Nan_tree}
\begin{equation}
\label{treefunc}
S_{k-\Delta k} [\phi_1, \phi_2] = \min_{\phi^{sp}_1, \phi^{sp}_2} 
\left(S_k[\phi_1 + \phi^{sp}_1, \phi_2 + \phi^{sp}_2] \right)\,.
\end{equation}
In case of two layers the exchange of the field variables is a symmetry of the model,
therefore the saddle point can be considered the same for both fields, namely:
\beq
\phi^{sp}_1 = \phi^{sp}_2 \equiv \phi^{sp} = 2 \rho_k \cos(k x^1).
\eeq
It is assumed that the periodic coordinate dependence appears only in
the space direction, which suits well to the treatment of the inhomogeneous
ground states. Restricting ourselves to a finite interval $[-L;L]$ there is a periodic boundary
condition for the saddle point, $\phi^{sp}(x-L)=\phi^{sp}(x+L)$.
Then the tree level blocking for the boson mass $M_k$ has the form
\beq
M_{k-\Delta k}(\phi_0^1-\phi_0^2)^2=M_k(\phi_0^1-\phi_0^2)^2\to k\partial_k M_k = 0.
\eeq
so $M_k\equiv M$ is a constant in the case of tree level blocking as well. Thus, $\tilde M_k\sim k^{-2}$
is a relevant coupling for all scales. The blocking step
\bea\label{blst}
V_{k-\Delta k} [\phi_1^0, \phi_2^0] &=& \min_{\rho_k} 
\biggl[2 \rho_k k^2+\sum_{n_1,n_2}\biggl[(u_{n_1n_2}+v_{n_1n_2})
\cos(\beta(n_1\phi_1^0-n_2\phi_2^0))(2(n_1-n_2)\beta \rho_k)\nn
&&~~~~~~+(u_{n_1n_2}-v_{n_1n_2})
\cos(\beta(n_1\phi_1^0+n_2\phi_2^0))(2(n_1+n_2)\beta \rho_k)\biggr]\biggr]
\eea
determines the tree level evolution for the couplings. The SG and the MSG models showed that
the phases of these models can be distinguished by the appearance of the spinodal instability
\cite{Nagy_SG,Nagy_MSG}. Since only the fundamental mode is relevant both in the SG and the MSG models,
perhaps it is not so surprising that the condition of spinodal instability gave a very good
approximation for a single coupling. Close to the scale of spinodal instability
$k_\mr{SI}$ the effective potential starts to form a parabolic shape \cite{Nagy_SG}, so the
minimum of the blocked potential is situated always at the values of the field variables
$\phi_1=\phi_2=0$. Taking into account the first few couplings ($\t u_{01}$, $\t u_{11}$ and
$\t v_{11}$) and inserting it into \eqn{sicond} one obtains
\bea
(1+2\t M_k-\t u_{01}\beta^2-\t u_{11}\beta^2-\t v_{11}\beta^2)
(1-\t u_{01}\beta^2-\t u_{11}\beta^2+\t v_{11}\beta^2) &=&0.
\label{silsg}
\eea
The fundamental mode follows the scaling relation
\beq\label{u01UV}
\t u_{01} = \t u_{01}(\Lambda)\left(\frac{k^2+2M}{\Lambda^2+2M}\right)^{\beta^2/16\pi}
\left(\frac{k}{\Lambda}\right)^{-2+\beta^2/8\pi}.
\eeq
The scale $k_\mr{SI}$ can be situated below or above the mass scale $M$, so one can
distinguish two cases:
\begin{enumerate}
\item $k_\mr{SI}<M$. The results of \app{app:ir16} show $\t u_{11}=\t v_{11}$ (the massive modes)
and \eqn{silsg} reduces to
\beq
(1+2\t M_k-\t u_{01}\beta^2-2\t u_{11}\beta^2)
(1-\t u_{01}\beta^2) = 0.
\label{silsg2}
\eeq
The expression cannot be zero if $\beta^2>\beta_c^2$ since $\t u_{01}$ is 
irrelevant. If $\beta^2<\beta_c^2$ then the expression $1-\t u_{01}\beta^2$ 
always go below zero, because now $\t u_{01}$ is relevant and grows up.
\item $k_\mr{SI}>M$. Then the massive modes are irrelevant keeping their
UV scaling according to \eqn{u01UV} but the coupling $\t u_{01}$ scales
relevantly when $\beta^2<\beta_c^2$, which implies that the expression in 
the second parenthesis in \eqn{silsg} can be negative. When 
$\beta^2>\beta_c^2$, then there is no spinodal instability.
\end{enumerate}
The discussion gives that the critical value of $\beta$ changes as 
compared to the case of the SG model, now $\beta_c^2=16\pi$.
Since in our model $\beta^2 = 4\pi$ the spinodal instability always occurs. It was
shown \cite{Nandori_LSG} that in this phase the couplings $\t u_{n_1n_2}$ are
relevant at least in the UV region.

\begin{figure}[ht]
\begin{center}
\epsfig{file=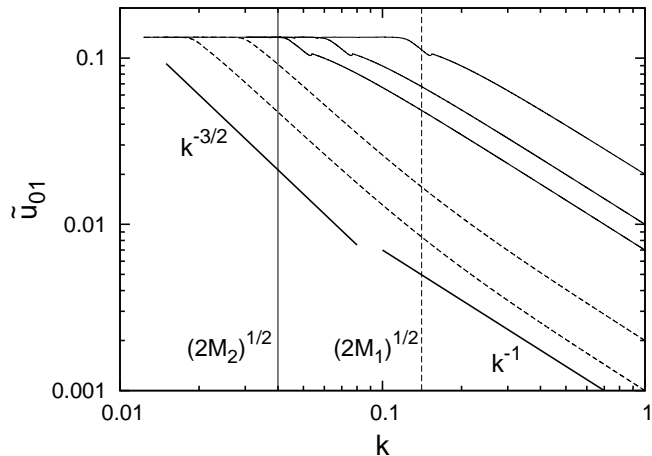,width=6cm,angle=-90}
\caption{\label{fig:irun}
The scaling of the fundamental mode $\t u_{01}$ for two different values bosonic mass,
$M_1=0.141$ (dashed lines) and $M_2=0.04$ (solid lines). For $M_1$
the UV ($k^{-1}$) and the IR ($k^{-3/2}$) scaling laws are drawn as reference.}
\end{center}
\end{figure}

We have taken into account 8 couplings when we solved the evolution equations
numerically. As in \cite{Nagy_SG,Nagy_MSG} we obtained that the higher modes do not affect
the scaling of the fundamental mode.
However the other couplings flow by different scaling behavior as obtained
by an extended UV RG approach \cite{Nandori_PLB,Nandori_LSG}. We started the evolution with the
WH-RG method then at the scale $k_{SI}$ we turned to the tree level blocking equations.
In the region $\beta^2>\beta_c^2$ a very interesting scaling law appears, as shown in \app{app:ir16}.
Furthermore, when $\beta^2=4\pi$ we run into the region of spinodal instability. The numerical
results can be seen in \fig{fig:irun}.
We numerically obtain that the dimensionless fundamental mode goes to a constant as in
\cite{Nagy_SG} independently of the relation between $k_{SI}$ and $M$, and the IR value
of $\t u_{01}(0)$ is superuniversal, independent of any UV parameter.
When $k_{SI}<M$ the flow of the fundamental mode qualitatively changes as the scale $k$ goes below $M$.
From \eqn{u01UV} one obtains $\t u_{01}(k\to\infty) \sim k^{-1}$ in the UV limit
but well below the mass scale $M$ when $k^2\ll 2M^2$ the flow scales according to
$\t u_{01}(k\ll M) \sim k^{-3/2}$, and after we reach the scale $k_{SI}$ and go towards
the the IR  regime the flow runs into $\t u_{01}(0)$. The evolution is similar
in the case of $k_{SI}>M$, but since the tree level blocking relation in \eqn{blst} is
independent of $M$, then below $M$ the scaling cannot change. Due to the necessity of the double Fourier
expansion we are unable to get as reliable numerical data as in \cite{Nagy_SG}, and we cannot
read off the IR form of the effective potential. From the solution of the condition $S''_k=0$
\cite{Nagy_SG,Nagy_MSG} and the qualitative value of $\t u_{01}(0)\approx 0.14$ one expects
\beq
\t V_{k\to 0} = -\hf(\phi_1^2+\phi_2^2).
\eeq
The IR physics of the zero density LSG model is free, the dimensionful coupling vanishes as $k\to 0$
at $\beta^2=4\pi$. However there are two phases with the critical value $\beta_c^2$
and they can be distinguished by the different scaling behavior of the dimensionful
couplings, namely when $\beta^2<\beta_c^2$, then the coupling scales as $u_{01}\sim k^2$, while
in the case of $\beta^2>\beta_c^2$ the scaling relation is $u_{01}\sim k^{\beta^2/8\pi}$.
Nevertheless according to the flow of the dimensionless couplings, in the phase $\beta^2<\beta_c^2$
the fundamental mode goes to a constant value, so it is marginal, but in the other phase it vanishes,
therefore it is irrelevant in the IR limit.

\subsubsection{Massless fermions in multi-flavor QED$_2$}
One can easily generalize the previous analysis for multi-flavor systems. We note
that the critical value is $\beta_c^2=8\pi N/(N-1)$ \cite{Nandori_PLB}, so
the spinodal instability always appears at $\beta^2=4\pi$.
The results of the RG analysis can be easily generalized and can show that the dimensionless
coupling $\tilde u_k$ tends to constant values in the IR limit.
Transforming the couplings of the scalar model ($u_k$ and $M_k$) into
the fermionic ones ($g_k$ and $m_k$) one has
\beq
M^2_k = \frac{e_k^2}{\pi}, \quad u_k = \frac{e^\gamma}2\pi^{-\frac{2N+1}{2N}}
m^\frac{2N-1}{N}e^\frac1{N}.
\eeq
The dimensionful minimal coupling $e_k$ does not scale independently of $N$, therefore
the fermionic mass $m_k$ should follow the irrelevant scaling of $u_k$
\beq
m_{k\to 0} \sim k^{2\frac{N}{2N-1}}\to 0,
\eeq
meaning that in the IR limit we always have massless multi-flavor QED$_2$. The pure QED$_2$ with $N=1$
is not massless since, the tree level blocking gives trivial (which is relevant in $d=2$)
scaling for the dimensionless coupling $\tilde u_k$ so the fermionic mass becomes
constant, showing that the cases $N=1$ and $N\ne 1$ significantly differs \cite{Nandori_PLB,Hosotani}.

\subsection{Low density phase}
Now let us turn into the case when $b\ne 0$.
The renormalization is rather involved in finite-density systems due to the
coordinate dependent ground state field configurations.
However in the low density phase average baryon density is zero (the low external baryon density
is totally screened by the induced density), and the resulting Hamiltonian
becomes the same as in the zero density case. Therefore we look for
the RG evolution of the model eventually among constant field configurations
just in the case of $b=0$, so all the results obtained there are valid in this phase too.
It implies that the low density LSG model is free, the fundamental mode is marginal, and
the dimensionful coupling scales as $u_{01}\sim k^2$. The corresponding
low density 2-flavor QED$_2$ is massless in the IR limit.

\subsection{High density phase}
We consider the RG evolution only for the high density limit. Here we always have a non-trivial
saddle point $\phi^{sp}$, and the quantum fluctuations takes place around this field configuration.
However during the blocking steps $\phi^{sp}$ also changes, which contributes
to the evolution of order $\ord{\hbar^0}$, therefore we concentrate on the
evolution only of the $\phi^{sp}$. Using the tree level blocking relation in \eqn{treefunc},
one obtains
\bea
S_{k-\Delta k}[\phi] &=&
u'_{k-\Delta k} \frac1{b}\sqrt{\frac2{\pi}}\cos(\sqrt{2\pi}\phi)\sin(\sqrt{2\pi} bL)\nn
&=& \int_x\left[\hf(\partial_1 \phi^{sp})^2+u'_k\cos(\sqrt{2\pi}(\phi+\phi^{sp}+bx))\right],
\eea
with $\phi$ a constant field configuration.
In high density limit the amplitude $A_1$ is small and can be considered perturbatively.
Furthermore at high densities one can retain only the fundamental mode.
Then the numerically determined form of the saddle point field configuration is
\beq\label{lsgtree}
\phi^{sp} = -A_1\sin(\sqrt{2\pi}(bx+\phi))+A_1\sin(\sqrt{2\pi}\phi),
\eeq
with $A_1=u'_k/\sqrt{2\pi}b^2$.
After identifying the corresponding functionals in \eqn{lsgtree} one obtains
\beq
u'_{k-\Delta k} = u'_k \left(1-\frac{u'^2_k}{4 b^4}\right)
\label{lbblocking}
\eeq
for the infinitesimal blocking step. The blocking relation in \eqn{lbblocking} clearly shows that
dimensionful coupling $u'_k$ decreases, and tends to zero so as the amplitude $A_1$. Since
the ratio $\Delta k/k$ is kept small in the WH-RG procedure, the blocking relation in \eqn{lbblocking}
is valid for the dimensionless couplings too, implying that the fundamental mode is irrelevant in
the high density phase. The quantum
fluctuations really wash out the wavy field configurations although at that price that in the IR
limit one obtains a free theory in the high density phase too, so the high density 2-flavor QED$_2$
is also massless.

\section{Conclusions}\label{sec:con}
We determined the phase structure of the bosonized version of the 2-flavor QED$_2$ and showed, that
in the tree level the induced baryon density totally
screens the low external baryon density $b$. In the case of high densities the screening
is only partial, furthermore a wavy induced baryon density is obtained.
We also performed the renormalization of bosonized model in both phases. When $b$ is zero
then during the evolution a non-trivial saddle point appears and the tree level blocking
gives a dimensionful periodic potential which flattens out, implying
massless 2-flavor QED$_2$ in the IR limit. It is also the
case when $b$ is small, since there is a total screening and the model becomes similar
to that of zero density one. The evolution of the bosonized model in the high density phase shows
that the amplitude $A$ of the periodic field configuration decreases
as the quantum fluctuations are integrated out step by step so as the fermionic mass, and
in the IR limit we obtain a massless theory with a trivial ground state.
The scaling of the dimensionless fundamental mode is marginal in the low density and
irrelevant in the high density phases in the bosonized model, therefore the phases found at the tree level,
survive the RG evolution and exist in the IR limit.

\section*{Acknowledgement}
The author acknowledges the Grant \"Oveges of the National Office for Research and Technology,
and the Grant of Universitas Foundation, Debrecen.

\appendix
\section{Irrelevant renormalizable operator in the LSG model}\label{app:ir16}
The evolution equation will generate the higher order Fourier modes, so one should take
them into account according to the ansatz in \eqn{lsgU}.
We introduce a truncation in the number of couplings, namely $n_1,n_2\le 2$. Then we have 8 apparently
independent Fourier modes. The dimensionless WH-RG equation is
\beq
(2+k\partial_k)\t U_k
=-\alpha\log\left[\left(1+2\t M_k\right)
+\left(1+\t M_k\right)(\t U_k^{11}+\t U_k^{22})
+\t U_k^{11}\t U_k^{22}+2\t U_k^{12}\t M_k -(\t U_k^{12})^2\right]
\eeq
Now one has to Fourier expand the equation to get the evolution equation
for the couplings. The resulting system of ordinary differential equation
is tackled by a numerical program. As we mentioned the RG-flow always runs into the
spinodal instability when $\beta^2<\beta_c^2$.
\begin{figure}[ht]
\includegraphics[width=8cm]{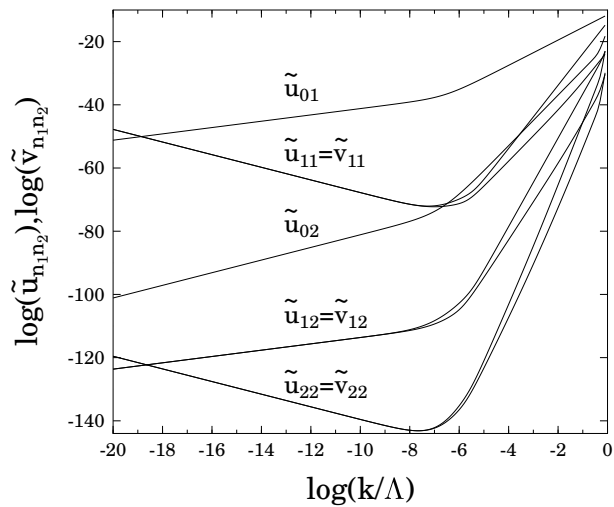}
\caption{For $\beta^2=24\pi$
the IR behavior of various Fourier amplitudes is presented for the LSG model.\label{irlsg}}
\end{figure}
In \fig{irlsg} one can see the typical IR flow of the couplings.
One can get the following conclusions from the numerical analysis:
\begin{enumerate}
\item
We numerically obtained that instead of 8 couplings we have only 5 independent one,
$\t u_{11}(k\to 0)=\t v_{11}(k\to 0)$, $\t u_{12}(k\to 0)=\t v_{12}(k\to 0)$ and
$\t u_{22}(k\to 0)=\t v_{22}(k\to 0)$, implying that the ansatz of the periodic part of
the potential in \eqn{lsgV} can be simply taken as
\beq
\label{lsg_ansatz2}
\t V_k=\hf M(\phi_1-\phi_2)^2+\sum_{n_1,n_2=0}^\infty
\t u_{n_1n_2}\cos(n_1\beta \phi_1-n_2\beta \phi_2).
\eeq
This result is independent of the value of $\beta$.
\item
At the scale around the mass of the theory $\sim \sqrt{\t M_k}$ the scaling of the
couplings change. There are modes which scales trivially below $\sqrt{\t M_k}$, namely
\beq\label{mas_sc}
\tilde u_{n_1n_1}\sim k^{-2},
\eeq
and they are referred as massive modes. Furthermore the other modes has qualitatively new scaling
behavior. Numerical results show that in general the Fourier amplitudes scales according to the law
\beq\label{scaling}
\tilde u_{n_1n_2}\sim k^{|n_1-n_2|(\frac{\beta^2}{8\pi}-2)-2\delta_{n_1,n_2}},
\eeq
with the Kronecker delta $\delta_{x,y}$. This result is also valid for arbitrary value of $\beta$,
but when $\beta<\beta_c$ the \eqns{mas_sc}{scaling} are correct until $k>k_{SI}$. Below
the scale of the spinodal instability the scaling relations of the flow changes.
\item
Starting from different UV initial values of the Fourier modes one obtains that only the
fundamental mode $\t u_{01}$ is sensitive for the initial conditions. The sensitivity matrix has the
same structure as in the case of the SG and MSG models \cite{Nagy_MSG,Nagy_SG} where the matrix has
only one nonzero column. We numerically obtained that the quantity
\beq
\frac{\t u_{n_1n_2}(k\to 0)}{\t u_{01}^{n_1+n_2}(\Lambda)} \equiv c_{n_1n_2}(k)
\eeq
is independent of $\t u_{n_1n_2}(\Lambda)$ and $\t v_{n_1n_2}(\Lambda)$. Consequently,
according to the sensitivity matrix the only relevant operator is the fundamental mode $\t u_{01}$ 
which drives the IR behavior of all the other couplings. On the other hand according to the 
scaling law in \eqn{scaling} $\t u_{01}\sim k^{\beta^2/8\pi-2}$ with 
positive exponent in the case of $\beta>\beta_c$ giving an irrelevant scaling.
\end{enumerate}
By the numerical solution of the RG equation derived for the LSG model we show
that for $\beta>\beta_c$ one can parameterize the RG flow of the couplings by the initial value 
of the fundamental mode $\t u_{01}(\Lambda)$, including the IR relevant massive modes.
The results of the sensitivity matrix shows that the fundamental mode is the only relevant 
operator of the model, nevertheless it goes to zero, so according to common classification of the couplings
it is irrelevant. Therefore, the scaling of all the higher harmonics and hence the low energy
effective theory of the LSG model is driven by an irrelevant coupling, showing the essential
importance of the investigation of irrelevant operators and the RG technique.

\end{document}